----------
X-Sun-Data-Type: default
X-Sun-Data-Description: default
X-Sun-Data-Name: talk
X-Sun-Content-Lines: 133

\documentstyle{article}

\begin{document}

\title{\bf MULTIPLICITY DISTRIBUTIONS IN QCD \\
AT VERY HIGH ENERGIES \\}

\author{\bf I.M. DREMIN \\}

\date{}
\maketitle
\begin{center}
\large{Lebedev Physical Institute,  Moscow,  Russia \\}
\end{center}
\vspace{3mm}
\begin{abstract}

It is shown that QCD is able to predict very tiny features of multiplicity
distributions at very high energies which demonstrate that the negative
binomial distribution (and, more generally speaking, any infinitely divisible
distribution) is inappropriate for precise description of experimental data.
New precise fits of high energy multiplicity distributions can be derived.

\end{abstract}

In this report, I briefly review the results of several cited below papers in
which it has been shown that it is possible to solve the strongly non-linear
integro-differential equations of QCD for the generating function of
multiplicity distributions. From the solutions obtained, it follows the
prediction about quite peculiar behavior of cumulants of the distribution which
was never found before. The accelerator data support the prediction. Thus, more
precise fits can be proposed to use in the cosmic ray range of very high
energies.

First, let us introduce some definitions related to the normalized multiplicity
distribution
\begin{equation}
P_n = \sigma _n /\sum _{n=0}^{\infty }\sigma _n
\end{equation}
of probabilities $P_n$ for $n$-particle (prong) events. Its generating function
$G(z)$ is defined as
\begin{equation}
G(z)=\sum _{n=0}^{\infty }(1+z)^{n} P_n  ,
\end{equation}
factorial moments are
\begin{equation}
F_q = \frac {\sum n(n-1)...(n-q+1)P_n}{(\sum nP_n )^q} = \frac {1}{\langle n
\rangle ^q} \frac {d^{q}G(z)}{dz^{q}}\vert _{z=0}
\end{equation}
and cumulants are
\begin{equation}
K_q = \frac {1}{\langle n \rangle ^{q}} \frac {d^{q}\log G(z)}{dz^{q}}\vert
_{z=0}.
\end{equation}
We use also their ratio
\begin{equation}
H_q = K_q /F_q.
\end{equation}

For the sake of simplicity, we consider here only a single gluon jet with the
total energy in its c.m.s. equal to $Q$ and introduce the evolution parameter
as $y=\ln (Q^2 /Q_{0}^{2})$, where $Q_{0}^{2}$=const. The generating function
satisfies the QCD equation [1,2]
\begin{equation}
G'(y)=\int _{0}^{1}dx (\frac {1}{x}-\Phi _{r}(x))\gamma _{0}^{2} [G(y+\ln
x)G(y+\ln (1-x))-G(y)],
\end{equation}
where $\Phi _{r}(x)=(1-x)(2-x(1-x))$ is the regular part of the
Altarelli-Parisi kernel and $\gamma _{0}^{2}=6\alpha _{S}/\pi $ ($\alpha _{S}$
is QCD coupling constant). It is the integro-differential non-linear equation
with shifted arguments in non-linear part and, therefore, it seems impossible
to find its solution. However, in a series of papers [3-7] it was shown that
such a solution exists in higher-order perturbative QCD for the running
coupling constant. Moreover, in the case of the fixed coupling constant, one
was able to get an exact solution of the equation [8,9] (both in gluodynamics
and in QCD with quarks and gluons). The decisive role in that progress is
played by the usage of formulae (3),(4) proposed in [4] and by notion of the
well-known relation between factorial moments and cumulants:
\begin{equation}
F_q = \sum _{m=0}^{q-1} C_{m}^{q-1}K_{q-m}F_m ,
\end{equation}
where $C_{m}^{q-1}$ are the binomial coefficients. Usage of (3),(4),(6) enables
one to get additional relation between $F_q$ and $K_q$ which together with (7)
provides the knowledge of any function $F_q, K_q, H_q$ and, therefore, of
multiplicity distribution $P_n$. I shall not delve into mathematical details
leaving room for physics discussion. One can learn them from cited papers
[3-10].

\vspace{1mm}
MAIN PHYSICS CONCLUSIONS ARE:
\vspace{1mm}

1. KNO-scaling is valid at very high energies.

2. The shape of the distribution is much narrower than in the lowest
perturbative approximation (called DLA).

3. The tail of the distribution falls off approximately as $\exp [-a(n/\langle
n \rangle )^{\mu }]$ with $\mu > 1$.

4. The cumulants $K_q$ acquire negative values while factorial moments are
always larger than 1.

5. Factorial moments are steadily increasing with their rank $q$ while
cumulants oscillate with increasing amplitude at very large $q$. (Integer
values of $q$ are considered. The extension to non-integer values was proposed
in [11].)

6. The ratio $H_q$ also oscillates with first minimum positioned at $q=5$ that
reveals a new expansion parameter of QCD.

7. This parameter shows that one should be careful in application of
perturbative QCD to multiparticle production.

8. The property 6) has been confronted to experiment and has found very good
support of it (see [12]).

9. The properties 4), 6) demonstrate that the negative binomial distribution
(so popular nowadays) is inappropriate for precise fits of multiplicity
distributions because its cumulants are always positive.

10. Moreover, all infinitely divisible distributions are inappropriate since
they have positive cumulants.

11. The cluster models with the Poissonian distribution of clusters are
inappropriate too.

The implications of these conclusions are to be studied yet. However, it is
remarkable that perturbative QCD becomes a powerful tool in describing the soft
processes when properly treated with higher-order terms taken into account.
Thus , the widely spread opinion that perturbative QCD is inapplicable to soft
processes should be reconsidered. The only (however, important) shortcoming is
the hadronization stage treated still either in the framework of the local
parton-hadron duality hypothesis or in Monte-Carlo simulations with various
assumptions. However, from above consideration it is clear that qualitative
features are well reproduced (and even predicted!) by QCD at the partonic
level.
The demonstrated in [3-10] possibility of the proper treatment of the partonic
stage provides some hope. In particular, the energy dependence of partonic
multiplicities [10] and the ratio of the partonic multiplicities in gluon and
quark jets [8, 10] give good chances for further studies of the hadronization
stage when compared with experimental data.

What concerns cosmic ray studies, it is not clear yet how the above findings
influence the fragmentation region multiplicities that is of upmost importance
for these investigations. It should be analyzed in combination with the
knowledge of rapidity spectra in hadron-hadron and hadron-nucleus collisions.

In conclusion, the non-linearity of QCD has important implications for soft
hadronic processes and it can be accurately treated to provide new predictions
for very high energy particle interactions.

\vspace{2mm}
\begin{center}
\large{Acknowledgement}
\end{center}
\vspace{2mm}

This work has been supported by the Russian fund for fundamental research
(grant 94-02-3815) and by Soros fund.

\newpage
\begin{center}
\large{\bf REFERENCES}
\end{center}
\vspace{3mm}

1. I.V.Andreev, {\it Quantum Chromodynamics and Hard Processes at High
Energies}, M., Nauka, 1981 (in Russian).

2. Yu.L.Dokshitzer, V.A.Khoze, A.H.Mueller, S.I.Troyan, {\it Basics of
Perturbative QCD}, Gif-sur-Yvette, Editions Frontieres, 1991.

3. Yu.L.Dokshitzer, Phys.Lett. B305 (1993) 295.

4. I.M.Dremin, Phys.Lett. B313 (1993) 209.

5. I.M.Dremin, Mod.Phys.Lett. A8 (1993) 2747.

6. I.M.Dremin, V.A.Nechitailo, JETP Lett. 58 (1993) 945.

7. I.M.Dremin, B.B.Levtchenko, V.A.Nechitailo, J.Nucl.Phys. 57 (1994) 477.

8. I.M.Dremin, R.C.Hwa, Phys.Lett. B324 (1994) 477.

9. I.M.Dremin, R.C.Hwa, Phys.Rev. D49 (1994) 5805.

10. I.M.Dremin, V.A.Nechitailo, Mod.Phys.Lett. A9 (1994) to be published.

11. I.M.Dremin, JETP Lett. 59 (1994) 501.

12. I.M.Dremin, V.Arena, G.Boca et al, Phys.Lett. B (1994) to be published.

\end{document}